\def\bpi{\vec\pi}

   \documentstyle[12pt,epsf]{article}
   
\newcommand{\nd}[1]{/\hspace{-0.5em} #1}
\begin{document}
   \begin{center}
       {\Large\bf The Large-$N_c$ Renormalization Group}
   \vskip 1.0cm
       {\bf NICHOLAS DOREY}\\
       {\it Physics Department, University of Wales Swansea, } \\
       {\it Singleton Park, Swansea, SA2 8PP, UK}
   \vskip .4cm
       and
   \vskip .4cm
       {\bf  MICHAEL P. MATTIS} \\
       {\it Theoretical Division, Los Alamos National Laboratory,} \\
        {\it Los Alamos, NM 87545, USA}
   \end{center}
   \vskip 1.5cm
\begin{abstract}
In this talk\footnote{Talk presented at the 1995 International Workshop on
Nuclear and Particle Physics in Seoul, Korea},
we review how effective theories of mesons and baryons become
exactly soluble in the large-$N_{c}$ limit. We start with a generic
hadron Lagrangian constrained only by certain well-known
large-$N_{c}$ selection rules. The bare vertices of the theory are dressed
by an infinite class of UV divergent Feynman diagrams at leading order
in $1/N_{c}$. We show how all these leading-order diagrams can be
summed exactly using semiclassical techniques. The saddle-point field
configuration is reminiscent of the chiral bag:
 hedgehog pions outside a sphere of radius
$\Lambda^{-1}$  ($\Lambda$ being the UV cutoff of the effective
theory) matched onto nucleon degrees of freedom for $r\le\Lambda^{-1}$.
 The effect of this pion cloud is to renormalize the bare nucleon
mass, nucleon-$\Delta$ hyperfine mass splitting,  and
Yukawa couplings of the theory.
The corresponding \it large-$N_c$ renormalization group equations \rm for
these parameters are presented, and solved
explicitly in a series of simple models. We explain under what
conditions  the Skyrmion emerges as a UV fixed-point of the RG
flow as $\Lambda\rightarrow\infty$.
\end{abstract}

{\bf Introduction.} The large-$N_{c}$ limit of QCD \cite{GT} is thought
to retain all the
important dynamical features of the realistic case, such as
confinement, chiral symmetry breaking and asymptotic freedom, while at
the same time offering considerable simplifications for the effective
theory of hadrons. The effective theory of mesons becomes
semiclassical in this limit and Witten \cite{EW} has argued that large-$N_{c}$
baryons should be identified with {\em chiral soliton} solutions of
the corresponding meson field equations. The Skyrme model
\cite{skyrme,ANW} provides the
simplest possible realization of this idea and yields a surprisingly
accurate picture of baryon properties. In addition there are several
model-independent features of the chiral soliton approach, such as the
$I=J$ rotor spectrum for the baryons, and the ``proportionality rule''
(see rules {\bf 2}-{\bf 4} below), which can be regarded as pure
large-$N_{c}$ predictions.

More recently, a seemingly orthogonal
 approach to large-$N_{c}$ baryons has been
promoted by Dashen, Jenkins and Manohar  \cite{DMJ}.
These authors start from an
effective Lagrangian with explicit fields for baryons as well as
mesons and then obtain powerful constraints on the allowed spectrum
and couplings of the theory by demanding self-consistency as
$N_{c}\rightarrow\infty$. Pleasingly, these consistency conditions
include all the model-independent
predictions of the chiral soliton approach mentioned above. Despite
this agreement, there appear to be several fundamental
differences between these two pictures of large-$N_{c}$
baryons. First, the nucleon is represented as a
pointlike spinor field in one and a semiclassical
extended object in the other. Second, in the effective Lagrangian
 picture, baryon number is
an ordinary $U(1)$ Noether charge while in the soliton
picture it is identified with a
{\em topological} charge which does not correspond to any symmetry of
the Lagrangian. And third, the number of free parameters is
 far greater in the effective Lagrangian approach; for example the
nucleon mass is arbitrary in this case but is a fixed function of the
mesonic couplings in the soliton approach, and likewise for the Yukawa
coupling(s).

And yet, since both approaches purport to describe
QCD in the low- and medium-energy regimes, must they not be equivalent
to one another?
In this talk, which is based on our recent papers \cite{DM1,DM2}, we
will explain precisely how this equivalence comes about---at least to
leading order in $1/N_c.$ (We are optimistic that the equivalence
continues to hold order by order in the $1/N_c$ expansion.) A cartoon
of our research program may be seen in Fig.~1. Starting
with a large-$N_c$-compatible effective Lagrangian, we will use
semiclassical techniques to demonstrate equivalence to a (quasi) chiral bag,
with the role of the bag radius being played by the inverse UV
cutoff $\Lambda^{-1}$ which regulates
the divergent meson-baryon Feynman diagrams in the effective theory.
It is then natural to ask, under what circumstances can the ``continuum
limit'' $\Lambda\rightarrow\infty$ be taken,  corresponding to
the limit of zero bag radius? The answer to this question is fairly
intricate, and is discussed in detail in Ref.~\cite{DM2}. Under
certain favorable circumstances, the UV limit exists---and is in
fact a Skyrmion/soliton model. We will review the formulation of the
so-called large-$N_c$ renormalization group, which is the tool for
exploring this question (for a comparison with the Cheshire Cat Principle,
see Ref.~\cite{DM2}). This talk, therefore, covers two of the three
arrows depicted in Fig.~1 (similar ideas are arrived at
 by Manohar \cite{mn}).\\

\centerline{\epsfysize=4.5in\epsffile{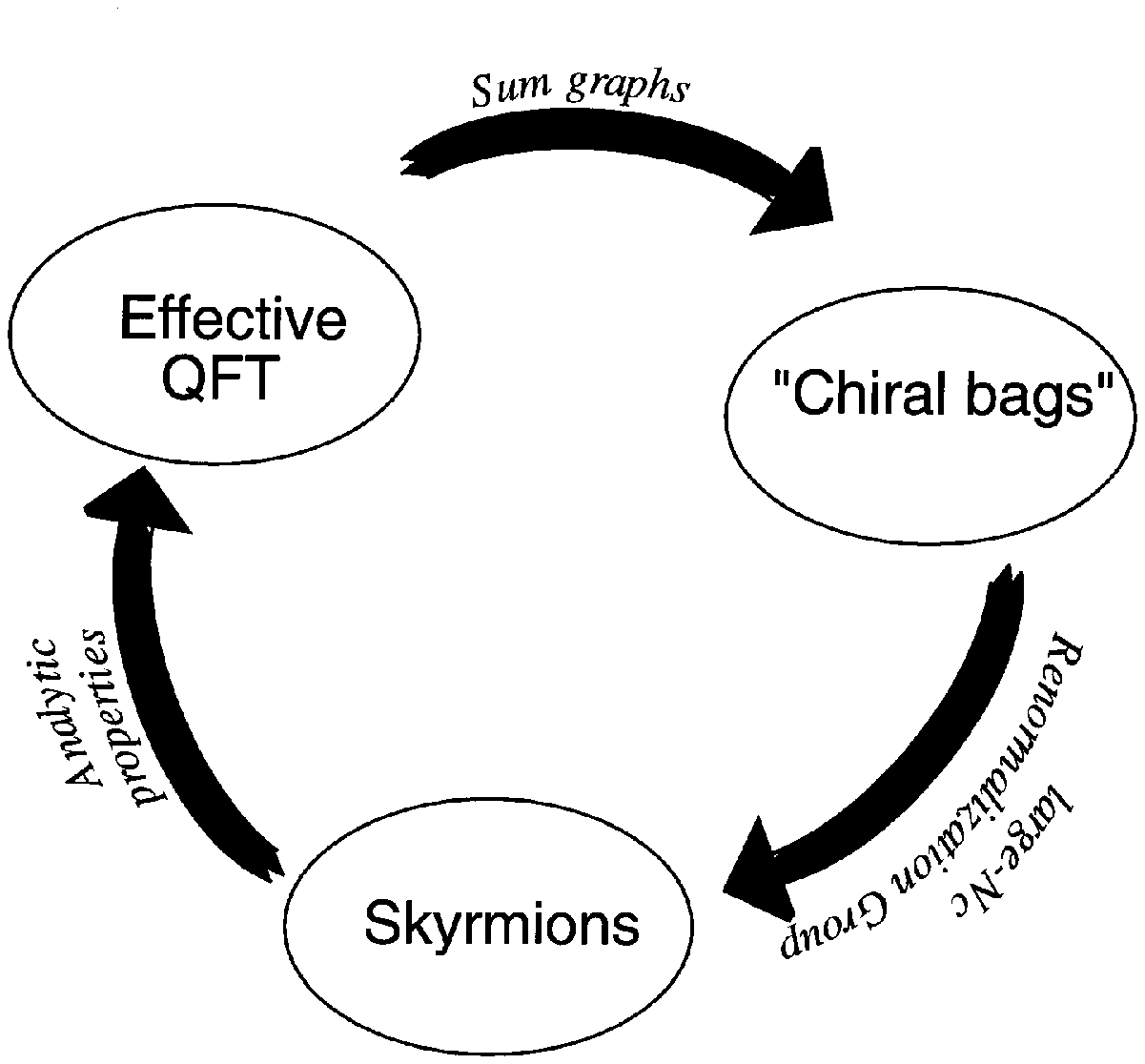}}\vskip4pt
\hangindent\parindent{{\bf Fig. 1}: Three types of large-$N_{c}$ models of the
 strong interactions, and the relationships between them.  This talk examines
 two of the three arrows, Effective QFT $\rightarrow$ ``Chiral bags,'' and
``Chiral bags'' $\rightarrow$ Skyrmions.  The third relation, Skyrmions
$\rightarrow$ Effective QFT, is examined in the following talk
\cite{DMRISKY}.}\\

\noindent The third arrow in Fig.~1 concerns how, starting
with a soliton model, one bootstraps one's way in the backwards
direction to an effective
Lagrangian; this is the topic of our talk which immediately follows
this one \cite{DMRISKY}.

{\bf Large-$N_c$ hadron models.} We study generic 2-flavor relativistic
hadron Lagrangians that conserve $C,$ $P,$ $T,$ and isospin,
and are further restricted \it only \rm by these
five large-$N_c$ consistency conditions:
\begin{description}
\item[1:] Straightforward quark-gluon counting arguments
show that $n$-meson vertices $\sim$  $N_c^{1-\frac{n}{2}}$,
 as do $n$-meson $2$-baryon vertices \cite{EW,LM,Georgi}.
Thus, baryon masses ($n=0$) and Yukawa couplings ($n=1$) grow like $N_c$
and $\sqrt{N_c}$, respectively.
\item[2:] The 2-flavor baryon spectrum of large-$N_c$ QCD consists of
an infinite tower of positive parity states with $I=J=1/2$, $3/2$,
$5/2\ldots$. To leading order
 these states are degenerate, with mass $M_{\rm bare}\sim N_{c}$
\cite{ANW,DMJ,LM,Georgi,GS}. (There are similar degeneracies amongst
the mesons that need not concern us here.)
\item[3:] Hyperfine baryon mass splittings
 have the form $J(J+1)/2{\cal I}_{\rm bare}$ where ${\cal I}_{\rm bare}
\sim N_c$ \cite{ANW,LM,Georgi,J1}.
\item[4:] Yukawa couplings are constrained to obey
the ``proportionality rule'' \cite{ANW,DMJ,GS,MatMuk,MatBraat}, which
fixes the interaction strength of
a given meson with each member of the baryon tower as a multiple of one
overall coupling constant (e.g., $g_{\pi N\Delta}/g_{\pi NN}=3/2$).
\item[5:] Finally, the allowed couplings of mesons to the baryon tower
must obey the $I_{t}=J_{t}$ rule \cite{MatMuk,MatBraat,MPM,Georgi}; e.g.,
the $\rho$ meson must be tensor-coupled to the nucleon
while the $\omega$ meson is vector-coupled at leading order in
$1/N_{c}$, in good agreement with phenomenology.
\end{description}
A concrete effective Lagrangian that embodies these selection rules,
and which is useful to keep in mind in the ensuing discussion,
is the following, consisting of
baryons and pions only:
\def\vpi{\vec\pi}
\begin{eqnarray}
{\cal L} & = & \frac{1}{2}(\partial_{\mu}\vpi)^2
-\frac{m_{\pi}^2}{2}\vpi^2 - V(\vpi)+
\bar{N}(i\nd{\partial}-M_{N})N \nonumber \\ & & \qquad{}
-g^{\rm bare}_{\pi}\partial_{\mu}\pi^{a}\bar{N}\gamma^{5}
\gamma^{\mu}\tau^{a}N + \hbox{(higher-spin baryons)}
\label{pionly}
\end{eqnarray}
Here $V$ is a general pion potential including quartic and higher vertices
subject to rule {\bf 1}; also
$M_{N}=M_{\rm bare}+3/8{\cal I}_{\rm bare}$
including the hyperfine splitting, as per rules {\bf 2}-{\bf 3}.
The pseudovector form of the $\pi N$ coupling is determined by the
$I_{t}=J_{t}$ rule while the proportionality rule fixes the
corresponding pion couplings to the higher-spin baryons.
The optional incorporation of additional meson species will be discussed below.

{\bf Summing the leading-order graphs.} Let us review how one
power counts effective meson-baryon Feynman graphs as per
the $1/N_c$ expansion. The $N_c$ dependence of
coupling constants described in rule {\bf 1} above trivially
suffices to identify
the leading-order meson-baryon Feynman graphs for any given physical process.
Thus, purely mesonic processes are dominated by meson
tree graphs, which vanish as $N_c\rightarrow\infty$; each meson
loop costs a factor of $1/N_c.$ Similarly,
meson-baryon processes are dominated by those graphs which become
meson trees if the baryon lines are removed.
\centerline{\epsfysize=6in \epsffile{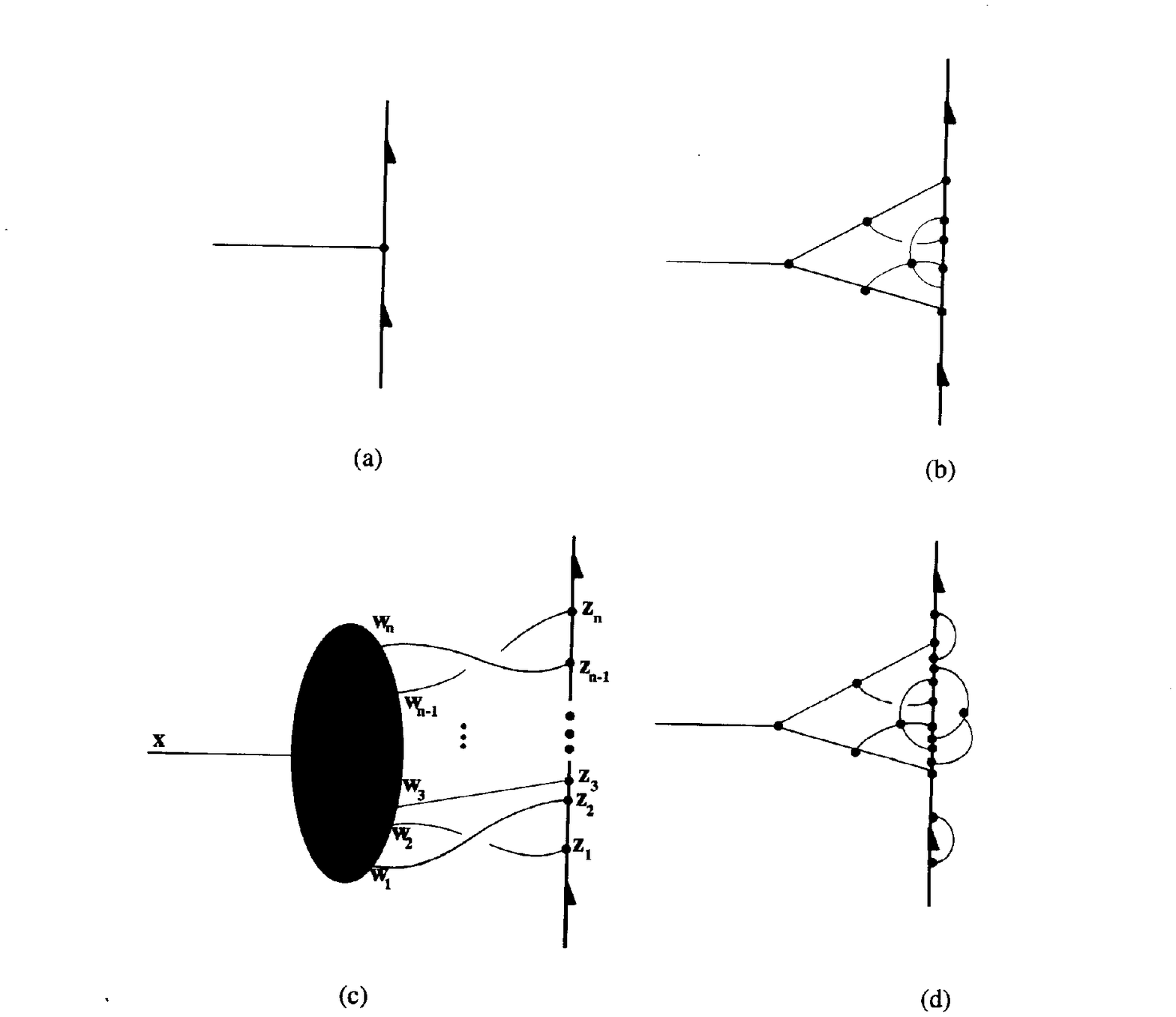}}\vskip4pt\hangindent
\parindent{{\bf Fig. 2}: (a) The bare meson-baryon coupling, which we shall
 refer to generically as a ``Yukawa coupling.''  Henceforth, directed lines
 are baryons, undirected lines are mesons.  Internal baryon lines must be
 summed over all allowed states in the $I = J$ tower.  (b) A typical
multi-loop dressing of (a) that contributes at {\it leading} order,
 $N_{c}^{1/2}$, as it contains no purely mesonic loops.  (c) A systematic
 counting of the diagrams such as (b).  The shaded blob contains only
 tree-level meson branchings.  There are $n!$ distinct ``tanglings'' of the
 attachments of the shaded blob to the baryon line.  (d) A typical dressing
 such as (b), augmented by additional baryon self-energy and vertex
corrections, all of which also contribute at leading order.}\\

To illustrate the uncontrolled
complexity of such graphs,  look at a typical multi-loop
correction, Fig.~2b, to the bare Yukawa coupling shown in Fig.~2a. Since,
by design, the graph in Fig.~2b contains no loops formed purely from
mesonic legs, this graph scales like $\sqrt{N_c}$ just like the bare
vertex. This is easily checked by multiplying together all the
vertex constants and ignoring propagators entirely, since both meson
and baryon propagators $\sim N_{c}^{0}$. Hence,
in order to calculate the dressed Yukawa vertex to leading
order, one must sum this infinite set of diagrams, shown somewhat
schematically in Fig.~2c. One must \it also \rm
sum all multiple
insertions of the baryon self-energy corrections and additional vertex
corrections as illustrated in Fig.~2d, as these too contribute at leading
order \cite{LM,J1}.
 Since many of the loop integrations in  these diagrams are UV
divergent, it is necessary to regulate the theory with a UV cutoff $\Lambda$.

It is clear that the naive perturbative graph-by-graph method is of
no use at all in large $N_c$ beyond the soft-pion regime; this is
because arbitrarily
complicated radiative corrections such as Fig.~2b-d contribute
to the renormalized Yukawa coupling at \it leading \rm order in the
$1/N_c$ expansion. In order to make progress, one had better sum
all such graphs, all at once. Pleasingly, such a summation
is in fact possible, if one restricts one's attention
to leading order in $1/N_c.$ This can be seen either
from momentum-space or position-space Feynman rules, following
Refs.~\cite{AM} or \cite{DM1}, respectively. The arguments are
a little more transparent in position space.
{}From the position-space Feynman rules for (\ref{pionly}),
the sum of all the graphs such as Fig.~2b (i.e., Fig.~2c)
is formally given by:
\def\calY{{\cal Y}}
\def\calA{{\cal A}}
\begin{equation}
\sum_{n=1}^{\infty}\int \left(\prod_{i=1}^{n} d^4w_{i}d^4z_{i}\right)
Blob_{n}(x;w_{1}, \ldots,w_{n}) \sum_{\rho \in S_{n}}\prod_{i=1}^{n}
g(w_{i},z_{\rho(i)})\cdot \calY(z_{\rho(i)})\prod_{i=1}^{n-1} G(z_{i},z_{i+1})
\label{formal}
\end{equation}
$Blob_{n}$ denotes the shaded blob in Fig.~2c; it contains the complete set of
tree-level meson branchings only, with no purely mesonic loops. The sum over
permutations $\rho \in S_{n}$ counts the $n!$ possible tanglings of the $n$
meson lines which connect the baryon line to the blob. All isospin and
spin indices have been suppressed in (\ref{formal}); intermediate baryons are
assumed to be summed over all allowed values of $I=J$.
$g_{ab}$ and $G_{J}$ are the
position-space meson  and spin-$J$ baryon propagators, respectively. Finally
$\calY_{JJ'}^a$ is the appropriate pseudovector Yukawa vertex factor connecting
the meson to an incoming (iso)spin $J$ and outgoing (iso)spin $J'$
baryon, according to the proportionality rule {\bf4} listed earlier.

Passing to the large-$N_c$ limit, we can exploit two important simplifications
to this expression. First, the baryons become very massive and can be
treated nonrelativistically. For forward time-ordering, $z_{0}<z'_{0}$, the
baryon propagator $G(z,z')$ can be replaced by its
nonrelativistic counterpart $G_{NR}(z,z')+{\cal O}(1/N_{c})$. As usual,
the reversed or ``$Z$-graph'' time ordering $z'_{0}<z_{0}$
contributes effective pointlike vertices in which two or more mesons couple to
the baryon at a single point. However these effective vertices turn out
to be down by two additional factors of $1/N_c$ since they are
proportional to $\big(\vec\sigma\cdot{\bf p}/M_N\big)^2$,
and we may neglect them.

A second simplification comes from anticipating the ``moral of our story,''
namely the equivalence to Skyrmion models, and borrowing from the
Skyrme-model literature a useful construct, namely the
$SU(2)$ collective coordinate basis for the $I=J$ baryons
 \cite{ANW,MatBraat,M}.
This is the basis denoted $|A\rangle$,
 related to the more familiar spin-isospin basis
$|I=J,i_{z},s_{z}\rangle$ via the overlap formula
\begin{equation}
\langle I=J, i_{z}, s_{z}|A\rangle = ({2J+1})^{1/2}D_{-s_{z},i_{z}}^{(J)}
(A^\dagger)\cdot(-)^{J-s_{z}}
\end{equation}
with $D^{(J)}(A)$  a  standard Wigner $D$-matrix.
In the $|A\rangle$ basis, the baryon propagator can be expressed as a \it
quantum mechanical \rm path integral over two collective coordinates:
${\bf X}$,  representing the position of the center of the baryon, and $A$,
describing its $SU(2)$ (iso)orientation,
\begin{eqnarray}
G_{NR}^{A,A'}(z,z') & = & \theta(z'_{0}-z^{}_{0})
\int_{{\bf X}(z_{0})={\bf z}}
^{{\bf X}(z'_{0})={\bf z}'}{\cal D}{\bf X}(t)\int_{A(z_{0})=A}
^{A(z'_{0})=A'}{\cal D}A(t) \nonumber
\\
& &  \qquad{} \times \exp{\left(i\int_{z_{0}}^{z'_{0}}dt\,
M_{\rm bare}+\frac{1}{2}M_{\rm bare}\dot{{\bf X}}^{2}
+{\cal I}_{\rm bare}{\rm Tr}\dot{A}^{\dagger}\dot{A}\right)}
\label{gnr}
\end{eqnarray}
The path integration over $A(t)$ can be performed using the beautiful
result of Schulman for free motion on the $SU(2)$ group manifold
\cite{schul},
\begin{eqnarray}
& &\int_{A(z_{0})=A}^{A(z'_{0})=A'}{\cal D}A(t)\,
\exp i\int_{z^{}_{0}}^{z'_{0}}dt\,
{\cal I}_{\rm bare}{\rm Tr}\dot{A}^{\dagger}\dot{A}\quad=
\nonumber
\\
& &\sum_{J=1/2,\,3/2,\cdots}\ \sum_{i_z,s_z=-J}^J
\langle A'|J,i_z,s_z\rangle\cdot
\exp i(z_0'-z^{}_0){J(J+1)\over2{\cal I}_{\rm bare}}\cdot
\langle J, i_{z}, s_{z}|A\rangle\ ,
\label{schu}
\end{eqnarray}
yielding the conventional nonrelativistic propagator for an infinite tower of
particles with masses $M_{\rm bare}(J)=M_{\rm bare}+
J(J+1)/2{\cal I}_{\rm bare}$ as required by the large-$N_c$ selection
rules.

For our present purposes, the greatest advantage of the $|A\rangle$ basis is
that the Yukawa vertex factor $\calY$ for the pion-baryon
coupling becomes \it diagonal \rm \cite{ANW,MatBraat,M}:
\begin{equation}
\calY_{A,A'}^{a}(z)=-3g^{\rm bare}_{\pi}D_{ab}^{(1)}(A)\delta(A-A')
\frac{\partial}{\partial z_{b}}
\end{equation}
Substituting for $G$ and $\calY$ in (\ref{formal}), we find that
this property allows us to perform the sum over
all tanglings of the meson lines trivially (the product over
temporal step-functions in Eq.~(\ref{gnr}) summing to unity).
Interchanging the order of path integration and the product over baryon
legs, one obtains
\begin{eqnarray}
& &\int {\cal D}{\bf X}(t){\cal D}A(t)\sum_{n=1}^{\infty}
 \int \left( \prod_{i=1}^{n} d^4w_{i}d^4z_{i}\right)
Blob_{n}(x;w_{1}, \ldots,w_{n})
\nonumber
\\
& &\times\ \prod_{i=1}^{n}
g(w_{i},z_{i})\cdot \calY(z_{i}) \delta^{(3)}_{\Lambda}
\big({\bf z}_{i}-{\bf X}(z^{0}_{i})\big)
\exp i S_{\rm baryon}[{\bf X},A]
\label{formal2}
\end{eqnarray}
where $S_{\rm baryon}$ is short for
the exponent of (\ref{gnr}). We have  reduced the problem to a
sum of \it tree \rm diagrams (namely, $Blob_n$)
for the pions interacting with the baryon
collective coordinates through a $\delta$-function source. The only remaining
manifestation of the UV cutoff $\Lambda$ is that this $\delta$-function
should be smeared out over a radius $\sim \Lambda^{-1}$, as
 denoted by $\delta_{\Lambda}$ in (\ref{formal2}), which we assume still
preserves rotational invariance.

In sum, the massive baryon has become a translating, (iso)rotating,
smeared point-source for the
pion field, the effect of which can be found be solving the
appropriate \it classical \rm Euler-Lagrange equation for a configuration
we call $\vec{\pi}_{\rm cl}(x;[{\bf X}],[A])$ \cite{AM}:
\def\sqr#1#2{{\vcenter{\vbox{\hrule height.#2pt
        \hbox{\vrule width.#2pt height#1pt \kern#1pt
           \vrule width.#2pt}
        \hrule height.#2pt}}}}
\def\square{\mathchoice\sqr84\sqr84\sqr53\sqr43}
\begin{equation}
(\square+m_{\pi}^{2})\pi_{\rm cl}^{a}+\frac{\partial
V}{\partial
\pi_{\rm cl}^{a}}=3g^{\rm bare}_{\pi}D_{ai}^{(1)}\big(A(t)\big)
\frac{\partial}{\partial x^{i}}
\delta^{(3)}_{\Lambda}({\bf x}-{\bf X}(t))
\label{ceq}
\end{equation}
It is easily checked (Fig.~3) that the order-by-order
perturbative solution of Eq.~(\ref{ceq})
 generates precisely the sum of
graphs appearing in (\ref{formal2}).\\

\centerline{\epsfysize=3.5in \epsfxsize=6in \epsffile{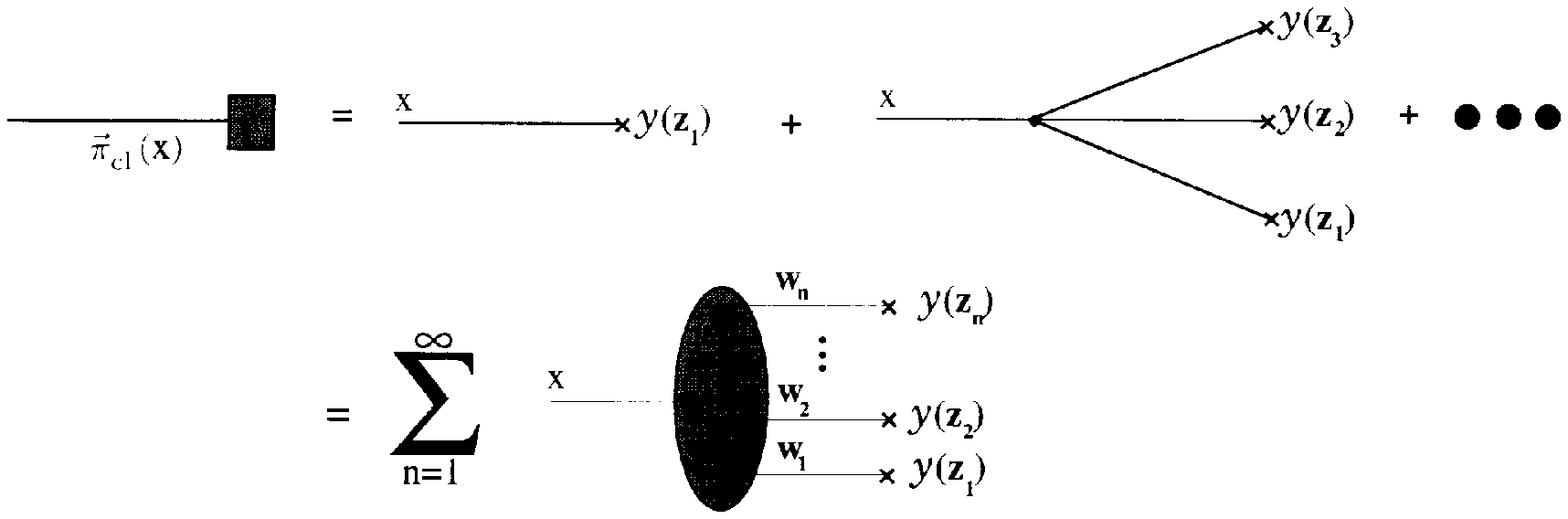}}\vskip-8pt
 \hangindent\parindent{{\bf Fig. 3}: The graphical perturbative solution to
 Eq.~(\ref{ceq})  as a sum of tree-level one-point functions terminating in
the effective Yukawa vertex.}\\

However, we still have not accounted for the baryon self-energy
and meson-baryon vertex corrections, highlighted in Fig.~2d. The key to
summing these is to notice that if the baryon line were erased, they
would be disconnected vacuum corrections, and that to leading order in
$1/N_c$ only the subset of such corrections that are meson trees are
important. As usual for vacuum corrections, they exponentiate. Furthermore,
by similar semi-classical reasoning as used above, they
are correctly accounted for by evaluating
the mesonic plus Yukawa pieces of the action (call this sum $S_{\rm eff}$)
on $\vpi_{\rm cl}$. This is illustrated in Fig.~4.\\
\centerline{\epsfysize=3.5in \epsffile{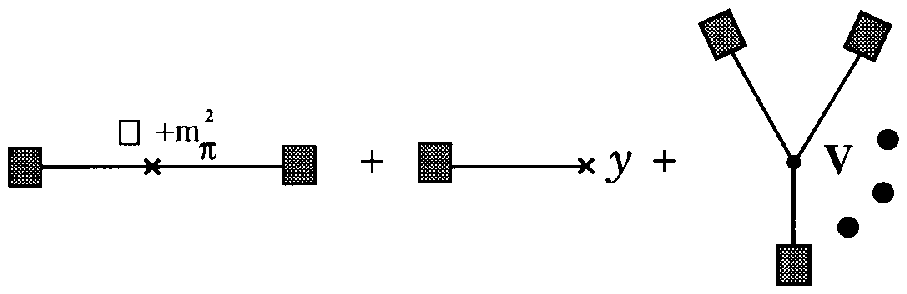}}\vskip4pt
 \hangindent\parindent{{\bf Fig. 4}: Diagrammatic representation of
$S_{\rm eff}(\vec{\pi}_{\rm cl})$.  When combined with the expansion depicted
 in Fig. 3, exp $iS_{\rm eff}$ combinatorically correctly accounts for all the
 leading-order baryon self-energy and vertex corrections highlighted in
 Fig. 2d.}\\

The final leading-order
result for the complete sum of graphs contributing to the dressed
pion-baryon vertex is then:
\begin{equation}
\int {\cal D}{\bf X}(t){\cal D}A(t)\, \pi^{a}_{\rm cl}(x;[{\bf X}],[A])
\exp i \big(
S_{\rm baryon}+ S_{\rm eff}[\vec{\pi}_{\rm cl},{\bf X}(t), A(t)]\,\big)
\label{ans}
\end{equation}
The blob has been eliminated from the problem, replaced simply
by $\vpi_{\rm cl}.$
For this to have happened, two conditions needed to be met. First,
the blob could be expressed as the complete sum of tree graphs, an
obvious consequence of large-$N_c$ since purely mesonic loops are suppressed by
$1/N_c.$ Second, and more subtlely,
there needed to exist a baryon basis (the $\left\langle A\right|$ basis)
in which the Yukawa source function ${\cal Y}$ is \it diagonal\rm.
The importance of this second ``diagonality'' condition was first emphasized by
Gervais and Sakita in their classic work on large $N_c$ \cite{GS}.
Only if \it both \rm conditions are met is a semiclassical summation
of the relevant Feynman graphs possible.

{\bf Solving the classical field equation.}
Now that we have reduced our problem to
 Eq.~(\ref{ceq}), one may ask, how does one actually \it solve \rm
such an equation? Again, we borrow Skyrme-model techniques,
 relating it to the analogous equation
for the \it static \rm pion cloud, $\vec{\pi}_{\rm stat}({\bf x})$,
surrounding a fixed baryon source (${\bf X}(t)\equiv{\bf0}, A(t)\equiv 1$):
\begin{equation}
(-\nabla^{2}+m_{\pi}^{2})\pi_{\rm stat}^{a}+\frac{\partial
V}{\partial\pi_{\rm stat}^{a}}=3g^{\rm bare}_{\pi}\frac{\partial}{
\partial x^{a}}\delta^{(3)}_{\Lambda}({\bf x})
\label{ceq2}
\end{equation}
The solution will generically have the hedgehog form
familiar from the Skyrme model:
$\pi^{a}_{\rm stat}({\bf x})=(f_{\pi}x^{a}/2r)F(r)$ where $r=|{\bf x}|$. The
profile function $F(r)$ is found, in turn,
by solving the induced nonlinear radial ODE.
While its detailed form depends sensitively on
the potential $V(\vpi)$,  its asymptotic behavior for
large $r$ is fixed by the linearized field equation,
\begin{eqnarray}
F(r) & \longrightarrow & \calA\left(\frac{m_{\pi}}{r}+\frac{1}{r^{2}}\right)
e^{-m_{\pi}r} \
\label{asm}
\end{eqnarray}
where the constant $\calA$ must, in the end, be extracted numerically.
The solution to  (\ref{ceq}) is then simply given, up
to $1/N_{c}$ corrections, by translating and (iso)rotating $\vpi_{\rm stat}$:
\begin{equation}
\pi_{\rm cl}^{a}(x;[{\bf X}],[A])=D^{(1)}_{ab}\big(A(t)\big)\pi_{\rm stat}^{b}
\big({\bf x}-{\bf X}(t)\big)
\end{equation}
The additional collective coordinate dependence carried by $\vpi_{\rm cl}$
versus $\vpi_{\rm stat}$ is precisely that required for overall isospin,
angular momentum and 4-momentum conservation, as is easily checked
\cite{DHM}.

We seek the renormalized on-shell $\pi N$ interaction, to leading
order in $1/N_c.$ It is defined in the usual way
as the on-shell residue of the LSZ amputation of the full
set of graphs that are summed implicitly by
 Eq.~(\ref{ans}). Formally, this amputation is
identical to the procedure one follows
in the Skyrme model \cite{DHM}. In particular, the
physically correct analytic structure of the one-point function
follows from the  $1/N_c$ corrections to $\vpi_{\rm cl}$ which describe
its response to the rotation of the source. (The specifics of this
response, involving an interesting small distortion \it away \rm
from the hedgehog ansatz \cite{DHM}, need not
concern us here; see the following talk \cite{DMRISKY}.)
 Thanks to the (iso)vector nature of the hedgehog,
the resulting S-matrix element for one-pion
 emission defines a renormalized on-shell pseudovector interaction of the pion
with the baryon tower, {\em identical} to the bare interaction in
 (\ref{pionly}), except for the coupling constant renormalization
 $g_{\pi}^{\rm bare}\rightarrow g_{\pi}^{\rm ren}.$ Again as in the
Skyrme model, this latter quantity is
determined by the asymptotics of $\vec{\pi}_{\rm stat}$, Eq.~(\ref{asm}),
and is explicitly given by \cite{ANW,DHM}
$g_{\pi}^{\rm ren}=(2/3)\pi f_{\pi}\calA$. Thus the proportionality and
$I_{t}=J_{t}$ rules for the pion-baryon coupling remain
true at the renormalized level, as claimed.

Furthermore, the result of evaluating $S_{\rm eff}[\vpi_{\rm cl}]$
is just an additive renormalization of the bare parameters of $S_{\rm baryon}$,
due to the meson cloud:
\begin{equation}
S_{\rm baryon}+S_{\rm eff}[\vec{\pi}_{\rm cl},{\bf X},A]\ =  \
\int dt\,\Big(
M_{\rm ren}+\frac{1}{2}M_{\rm ren}\dot{{\bf X}}^{2}
+{\cal I}_{\rm ren}{\rm Tr}\dot{A}^{\dagger}\dot{A}\,\Big)
\end{equation}
where
\begin{eqnarray}
M_{\rm ren}=M_{\rm bare}+\int d^{3}{\bf x}\,
({\nabla}\pi_{\rm cl}^{a})^2\ ,
& & {\cal I}_{\rm ren}={\cal I}_{\rm bare}+\frac{2}{3}\int d^{3}{\bf x}\,
\vpi_{\rm cl}^2
\end{eqnarray}
It follows that
$M_{\rm ren}(J)=M_{\rm ren}+J(J+1)/2{\cal I}_{\rm ren}$ and so the form
of the hyperfine baryon
mass splitting is likewise preserved by renormalization.
The \it self-consistency \rm of large-$N_c$ effective models, as evidenced by
the last two paragraphs, is one of the most striking features of
the large-$N_c$ approach; namely, that selection rules implemented
at the bare level survive the all-loops renormalization process.

The generalization of the above analysis to models including several
species of mesons  involves solving the
 coupled classical radial ODE's for all the meson fields, using
generalized hedgehog ansatze familiar from
vector-meson-augmented Skyrme models.
A particularly rich meson model might include, in addition to
the pion, the tensor-coupled
$\rho$, i.e., $g_\rho^{\rm bare}
\partial_\mu\vec\rho_\nu\cdot\bar N\sigma^{\mu\nu}
\vec\tau N,$ the vector-coupled $\omega,$ i.e.,
$g^{\rm bare}_\omega
\omega_\mu\bar N\gamma^\mu N,$ and/or the ``$\sigma$-meson,''
which couples simply as $g_\sigma^{\rm bare}
\sigma \bar N N.$ Again, on shell, the form
of these particular couplings survives renormalization.

{\bf Large-$N_{c}$ Renormalization Group.} We have
described an  explicit numerical procedure
for calculating the renormalized Yukawa couplings, baryon
masses and hyperfine mass splittings,
to leading order in $1/N_{c}$, directly from the classical meson cloud
surrounding the baryon. Since the $\delta$-function
source on the right-hand side of Eq.~(\ref{ceq}) is smeared out over a
characteristic length $\Lambda^{-1}$, these
quantities depend explicitly on $\Lambda$. In order to hold the
physical, renormalized masses and couplings fixed,
 it is necessary to vary simultaneously
{\em both} $\Lambda$ and the corresponding bare quantities.
This procedure defines an RG flow for
$M_{\rm bare}(\Lambda)$, ${\cal I}_{\rm bare}
(\Lambda)$ and $g_{\pi,\rho,\omega,\sigma}^{\rm bare}(\Lambda)$,
valid to all orders in the loop expansion but strictly to leading order in
$1/N_{c}$. We term this flow the \it large-$N_c$ Renormalization Group\rm,
and devote the rest of this talk to illustrating its solutions.

It is particularly interesting to ask whether this flow
has a UV fixed point; this would correspond to a continuum limit
for the theory. Unfortunately we are not able to prove any general
theorems about the RG flow for large-$N_{c}$ effective
theories. Nevertheless, in \cite{DM2}, we were able
to carry out the program outlined above
explicitly in a series of simple
but physically relevant models of pions with a pseudovector
coupling to the $I=J$ baryon tower. The models are distinguished from
one another only by the choice of the purely mesonic Lagrangian
denoted ${\cal L}_{\rm meson}$. For details we refer the interested
reader to \cite{DM2}, and here content ourselves with a summary of the
salient results.

Our first example consists simply of free massless pions,
\begin{equation}
{\cal L}_{\rm meson}=\frac{1}{2}(\partial_{\mu}\bpi)^{2}
\label{free}
\end{equation}
coupled derivatively to the $I=J$ baryon tower. In the
hedgehog ansatz, the static Euler-Lagrange equation (\ref{ceq2}) becomes
\begin{equation}
F''+\frac{2}{r}F'-\frac{2}{r^{2}}F=6f_{\pi}^{-1}g^{\rm bare}_{\pi}(\Lambda)
\frac{\partial}{\partial r} \delta_{\Lambda}(r)
\label{bageq}
\end{equation}
This being a linear equation, it
is trivially solved using the method of Green's functions:
\begin{equation}
F(r) =  6f_{\pi}^{-1}g_{\pi}^{\rm bare}(\Lambda)
\int_{0}^{\infty} dr'\,{r'}^2\,G(r,r')\,\frac{\partial}{\partial r'}
\delta_{\Lambda}(r')
\label{soln}
\end{equation}
where the Green's function that is well behaved at both $r=0$ and
$r=\infty$ is
\begin{equation}
G(r,r')=-\frac{r_{<}}{3r_{>}^{2}}
\label{gfn}
\end{equation}
where $r_{<}={\rm min}[r,r']$ and $r_{>}={\rm max}[r,r']$.

The renormalized
 Yukawa coupling $g_{\pi}^{\rm ren}$ is extracted from the large-distance
behavior of $F$ as per equation (\ref{asm}). With the mild (and relaxable)
assumption that $\delta_{\Lambda}(r')$ has compact support, Eqs
(\ref{soln}) and
and (\ref{gfn}) imply
\begin{equation}
F(r) \rightarrow \frac{3g_{\pi}^{\rm bare}(\Lambda)}{2\pi f_{\pi} r^2}
\label{extract}
\end{equation}
Comparing (\ref{extract}) and (\ref{asm}), we deduce
\begin{equation}
g_{\pi}^{\rm bare}(\Lambda)=g_{\pi}^{\rm ren}
\label{nonren}
\end{equation}
for all $\Lambda$,
admittedly not a surprising result for free field theory, but a reassuring
sanity check on our formalism. The simplest modification to the
Lagrangian (\ref{free})
is to add a pion mass term. In that case a similar analysis yields;
\begin{equation}
g_{\pi}^{\rm bare}(\Lambda)=g_{\pi}^{\rm ren}\cdot\left(1+
O(m_{\pi}^{2}/\Lambda^{2})\right)
\label{nonrenII}
\end{equation}
In either variation, massless or massive,
the ``continuum limit'' $\Lambda\rightarrow\infty$ can be safely taken,
and the ``ultraviolet fixed point'' that emerges is just what one started
with: a theory of free pions derivatively coupled to the baryon tower.

For our second example, consider the nonlinear $\sigma$ model for
pions,
\begin{equation}
{\cal L}_{\rm meson}=\frac{f_{\pi}^{2}}{16}{\rm Tr}\,\partial_{\mu}U
\partial^{\mu}U^{\dagger}
\label{sigma}
\end{equation}
where $U=\exp(2i\bpi\cdot\vec\tau/f_{\pi})$
again augmented by the bare pseudovector Yukawa coupling. The static
Euler-Lagrange equation now works out to
\begin{equation}
F''+\frac{2}{r}F'-\frac{1}{r^{2}}\sin 2F
=6f_{\pi}^{-1}g^{\rm bare}_{\pi}(\Lambda)
\frac{\partial}{\partial r} \delta_{\Lambda}(r)
\label{bageqII}
\end{equation}
Solving this nonlinear equation for $F(r)$
requires that we specify a smearing of the source.
For convenience, we follow \cite{mn}, and choose a radial step-function
\begin{equation}
\delta_{\Lambda}(r)=\frac{3\Lambda^{3}}{4\pi} \theta(\Lambda^{-1}-r)
\label{smear}
\end{equation}
which is properly normalized to unit volume. The technical advantage,
which we exploit presently, is that
the right-hand side of (\ref{bageqII}) is now proportional to a \it
true \rm $\delta$-function, since
\begin{equation}
\frac{\partial}{\partial
r}\delta_{\Lambda}(r)=-\frac{3\Lambda^{3}}{4\pi}
\delta(r-\Lambda^{-1})
\label{truedelta}
\end{equation}
There is also a conceptual advantage: the right-hand side of
(\ref{truedelta})
means that the baryon and meson degrees of freedom only interact
at the ``bag radius'' $\Lambda^{-1}$ which sharpens the analogy to the
traditional chiral bag (a topic we shall return to at the end of this talk).

With this convenient choice of regulator,
 the prescription for satisfying (\ref{bageqII}) is
transparent: First solve the \it homogeneous \rm version of equation
(\ref{bageqII})
for $r<\Lambda^{-1}$ (``region I'') and for $r>\Lambda^{-1}$
(``region II''); next,
match the solutions in these two regions,
$F_{\rm I}(r)$ and $F_{\rm II}(r)$, at the point $r=\Lambda^{-1}$;
and finally, read off $g_{\pi}^{\rm bare}(\Lambda)$ from the slope
discontinuity,
\begin{equation}
g_{\pi}^{\rm bare}(\Lambda)=\frac{2}{9}\pi
f_{\pi}\Lambda^{-3}\left(F_{\rm I}'(\Lambda^{-1})-F_{\rm II}'
(\Lambda^{-1})\right)
\label{disc}
\end{equation}
In Ref.~\cite{DM2} we implemented this prescription numerically and
determined the behaviour of $g_{\pi}^{\rm bare}$ as $\Lambda$ is
varied with $g^{\rm ren}_{\pi}$ held fixed at its experimental value.
We discovered a critical value of the cutoff, $\Lambda_{c} \simeq 340$
MeV above which there is no real solution, corresponding to a critical
bag radius $\Lambda_c^{-1}\simeq.6\,$fm. In this case, therefore,
there exists an obstacle to taking a continuum limit $\Lambda
\rightarrow \infty$.

The (little-known!) fact that a non-linear radial ODE in three dimensions
has no solution with a point-like $\delta$-function source strongly
suggests that any continuum limit of the RG flow
necessarily involves the coupling $g_{\pi}^{\rm bare}$, which
multiplies the source, tending to zero as $\Lambda\rightarrow\infty$.
In this case the resulting fixed-point
field configuration should be a solution of the {\em homogeneous} meson
field equation (ie with the source term set to zero). It is therefore
natural to conjecture that the model only has a continuum limit if the
homogeneous meson field equation admits a {\em chiral soliton}
solution. In this light, the occurrence of a critical cutoff for the
non-linear $\sigma$-model coupled to large-$N_{c}$ baryons has an
obvious explanation: the mesonic sector of this model, which consists
of a single two-derivative term, does not support a soliton solution
because of Derrick's theorem \cite{derrick}. Hence there can be no
continuum limit.

The simplest way to remedy this problem is to augment the mesonic
sector of the model by adding the Skyrme term. In this case ${\cal
L}_{\rm meson}$ is just the Lagrangian density of the standard two-term
Skyrme model.
\begin{equation}
{\cal L}_{\rm meson}=\frac{f^{2}_{\pi}}{16}{\rm Tr}\,\partial_{\mu}U^{\dagger}
\partial^{\mu}U+\frac{1}{32e^{2}}{\rm Tr}[U^{\dagger}
\partial_{\mu}U\, , \, U^{\dagger}\partial_{\nu}U]^{2}
\label{skyrme}
\end{equation}
In this case the radial ODE becomes
\begin{equation}
F''+\frac{2}{r}F'  -  \frac{\sin 2F}{r^{2}}\left(1+\frac{4\sin^{2}
F}{e^{2}f_{\pi}^{2}r^{2}}-\frac{4F'^{2}}{e^{2}f_{\pi}^{2}}\right)
 = 6f_{\pi}^{-1}g^{\rm bare}_{\pi}(\Lambda)
\frac{\partial}{\partial r} \delta_{\Lambda}(r)
\label{bageqIII}
\end{equation}
while the derivative matching condition (\ref{disc}) remains
unchanged. We know that the homogeneous pion field equation obtained
by setting the RHS of (\ref{bageqIII}) to zero has a soliton solution;
the Skyrmion. Hence a naive conjecture is that the Skyrmion arises as
the UV fixed point of the RG flow in these models. However, as stated,
this naive conjecture cannot be right! The reason is the mismatch
in the number
of free parameters between soliton/Skyrmion theories and effective
Lagrangian theories, mentioned earlier. In particular, in the
latter the Yukawa coupling is independent of the mesonic parameters,
whereas in the Skyrme model
the asymptotic behaviour of the Skyrmion at large
$r$ fixes a unique value for $g_{\pi}^{\rm ren}$ originally
determined numerically by Adkins, Nappi and Witten \cite{ANW};
\begin{equation}
g_{\pi}^{\rm ren}\equiv g_{\pi}^{\rm ANW}\simeq
\frac{18.0}{e^{2}f_{\pi}}
\label{ganw}
\end{equation}
 Our analysis \cite{DM2}  of the
equation (\ref{bageqIII})  resolves this conundrum: it
reveals that the Skyrmion emerges as a UV
fixed point {\em only} when $g_{\pi}^{\rm ren}$ is fine-tuned to obey
(\ref{ganw}). If $g_{\pi}^{\rm ren}\neq g_{\pi}^{\rm ANW}$ then there
again exists a critical value of the UV cutoff beyond which the RG
flow does not go. The obstacle in this case is somewhat different to
that which occurs when only the two-derivative term is included. There
the solution of equation (\ref{bageqII})
actually ceases to exist above the critical cutoff. Here the solution
still exists but is no longer locally stable and is thus no longer an
appropriate saddle-point. We refer the reader to \cite{DM2} for further
details.

In summary we have shown how a generic, large-$N_{c}$ consistent,
effective theory of hadrons can be solved exactly as
$N_{c}\rightarrow\infty$
using semi-classical methods. We should stress that this solubility is
precisely a consequence of the large-$N_{c}$ selection rules
incorporated in the bare Lagrangian; for
example, without the proportionality rule, which implies that the
baryon-pion vertex is diagonal in the $A$ basis, it would {\em not} have been
possible to sum over all possible spins and isospins of the
internal baryon lines with a single saddle-point field configuration.
Correspondingly, the hedgehog structure of the
resulting pion cloud ensures that the large-$N_{c}$ selection rules
emerge unscathed at the renormalized level; an important check on the
self-consistency of large-$N_{c}$ effective theories.

We also showed how the large-$N_c$
RG equations for the flow of the bare
Lagrangian parameters can be determined exactly at leading order, and we
exhibited their solutions in a series of specific models. Our
results for the Skyrme Lagrangian {\em coupled to explicit baryon
fields} show how the Skyrmion emerges naturally as the continuum limit of the
dressed large-$N_{c}$ baryon. From our construction, it is easy to see
how the apparant differences, discussed in the Introduction,
between the effective Lagrangian and
chiral soliton treatment of large-$N_{c}$ baryons are
resolved. Although the bare nucleon in the former approach is a
point-like field, it is dressed at leading order by an infinite set of
Feynman diagrams which sum up to give a semiclassical pion cloud which
coincides with the Skyrmion at large distances. As the continuum limit
is taken the explicit baryon number carried by the bare nucleon is
completely screened by the topological charge of the Skyrmion in a manner
familiar from the chiral bag model \cite{GJ}. Finally the puzzle about the
number of free parameters in the two approaches has an obvious
solution; the continuum limit only exists when the renormalized
baryon parameters are fine tuned to obey Skyrme-model relations. In the
language of the renormalization group, the corresponding terms in the
bare Lagrangian (e.g., the Yukawa couplings) are {\em irrelevant
operators}. In contrast, the mesonic
self-couplings in the Lagrangian which determine the soliton solution
are not renormalized at leading order in $1/N_{c}$
because purely mesonic loops are subleading. Hence they
correspond to the {\em marginal operators} of the large-$N_{c}$
renormalization group which dominate in the continuum limit.

 In this manner,
the large-$N_c$ Renormalization Group in the UV limit $\Lambda\rightarrow
\infty$ serves as the long-sought connection
between effective Lagrangians and and Skyrmion
models of the baryon (Fig.~1 again).
But as alluded to earlier,
for fixed, finite $\Lambda,$ the picture we have arrived at of the
meson-dressed large-$N_c$ baryon is highly reminiscent of yet a third
class of phenomenological models, the chiral bag
models \cite{GJ}. These, too, are
hybrid descriptions of the dressed baryon, in which explicit quark
(rather than nucleon) degrees of freedom inside a bag of radius
$R=\Lambda^{-1}$ are  matched onto an effective theory of hedgehog
pions outside the bag. Even this presumably important distinction
between `nucleon' versus `quark' degrees of freedom inside the bag
disappears as $N_c\rightarrow\infty.$\footnote{This observation was originally
made by Witten (Ref.~\cite{EW}, Secs.~5 and 9), and exploited by Gervais and
Sakita (Ref.~\cite{GS}, Sec.~V). These two papers are highly recommended
background reading, as they too are concerned primarily with
 the semiclassical nature of large-$N_c$. In particular Gervais and
Sakita were the first to study chiral-bag-type structures in this
limit, although not from our effective hadron Lagrangian starting point.}
 For, in this limit, the $N_c$
quarks may be treated in Hartree approximation, and their individual
wave functions effectively
condense into a common mean-field wave function, which we
may identify with the ``wave function of the nucleon.'' Outside
the bag, the analogy is closer still: the pion field
configuration is again determined by solving a nonlinear field equation
coupled to a static source at $r=R$.  The only significant difference
between our composite meson-dressed large-$N_c$ baryon and the traditional
chiral bag is this: our composite baryon follows solely from large $N_c$
and has \it nothing whatsoever
to do with chiral symmetry! \rm It is for this reason that we  referred to
it as a ``chiral bag'' in Fig.~1,  being careful to retain the quotation
marks to avoid confusion with the traditional chiral bag.

 The link in the opposite direction, from Skyrmions
to effective Lagrangians, is presented in the talk to follow \cite{DMRISKY}.

\end{document}